\documentclass[prl,showpacs,preprintnumbers,amsmath,amssymb,letterpaper]{revtex4}
\usepackage{graphicx}
\usepackage{dcolumn}
\usepackage{bm}
\usepackage{epstopdf}
\usepackage{latexsym}
\usepackage{epsfig}
\usepackage{verbatim}
\usepackage{hyperref}
\usepackage[usenames]{color}

\begin{document}
\title{Saturation of the Anomalous Hall Effect in\\Critically
Disordered Ultra-thin CNi$_3$ Films}
\author{Y. M. Xiong, P. W. Adams}
\affiliation{Department of Physics and Astronomy, Louisiana State
University, Baton Rouge, LA 70803-4001, USA}
\author{G. Catelani}
\affiliation{Department of Physics, Yale University, 217 Prospect
Street, New Haven, CT 06520, USA}
\date{\today}

\begin{abstract}
We demonstrate that a distinct {\it high-disorder} anomalous Hall effect phase emerges at the correlated insulator threshold of ultra-thin, amorphous, ferromagnetic CNi$_3$ films.  In the weak localization regime, where the sheet conductance $G\gg e^2/h$,  the anomalous Hall resistance of the films increases with increasing disorder and the Hall conductance scales as $G_{xy}\propto G^\varphi$ with $\varphi=1.6$.  However, at sufficiently high disorder the system begins to enter the 2D correlated insulator regime, at which point the Hall resistance $R_{xy}$ abruptly saturates and the scaling exponent becomes $\varphi=2$.  Tunneling measurements show that the saturation behavior is commensurate with the emergence of the 2D Coulomb gap, suggesting that $e$-$e$ interactions mediate the high-disorder phase.
\end{abstract}


\maketitle

One of the most far-reaching goals in condensed matter physics is to develop a comprehensive understanding of how electron-electron ($e$-$e$) interactions, and their attendant correlations, affect the electronic properties of two dimensional (2D) systems.    It is now well known that reduced dimensionality often induces profound changes in the electronic behavior of materials, particularly in the presence of disorder. Indeed, interesting manifestations of electron correlation effects have been reported in a wide variety of 2D systems, including homogeneously disordered metal films \cite{Lee,Altshuler}, thin film superconductors \cite{S-I}, and two dimensional electron gasses in semiconducting heterojunctions \cite{Bishop}, to name a few.   In the present Letter, we show that the anomalous Hall effect (AHE)  \cite{Nagaosa} also offers a compelling stage from which to study the manifestations of disorder-enhanced quantum correlations but from a perspective in which spin-dependent scattering processes are important.  In addition, because the effect arises from both magnetic scattering and intrinsic band mechanisms, it remains a topic of current interest, with relevance to the burgeoning field of spintronics.

The AHE is characterized by the appearance
of a spontaneous Hall resistance in magnetic materials and is usually
parameterized by the empirical relation \cite{Bergmann,Hurd,Bergmann2},
\begin{equation}
\label{Hall Effect}
R_{H} =V_H/I= (R_{o}B + R_{e}\mu_{0}M)/t,
\end{equation}
where $R_{H}$ is the Hall resistance, $V_H$ is the Hall voltage, $I$ is the longitudinal
current, $R_{o}$ is the ordinary Hall coefficient, $R_{e}$ is the
anomalous (extraordinary) Hall coefficient, $M$ is the
magnetization, and $t$ is the film thickness. The term proportional to $M$
in Eq.~\ref{Hall Effect} produces the anomalous Hall voltage and, in
transition metal ferromagnetic films, it is usually orders of
magnitude larger than the ordinary Hall term.  The robustness of the AHE signal offers one a potentially important probe of localization effects as manifest in the three predominant AHE channels: {\it intrinsic inter-band scattering}, {\it skew scattering},
and {\it side-jump scattering} \cite{Nagaosa}.  The theoretical treatment of the AHE has a long and contentious history,
but in recent years there has been a significant interest in how disorder-enhanced quantum correlations affect
the low-temperature behavior of the AHE and, in particular, how
such corrections are related to the well-established weak-localization
corrections to the longitudinal transport \cite{Lee}. This remains
an open issue, particularly in the regime of moderate to strong disorder \cite{Bergmann,Langenfeld,Lin,Mitra}.
In this Letter, we present a study of the low-temperature
anomalous Hall effect as a function of the sheet resistance of
homogeneously disordered CNi$_3$ films.   Using magnetotransport and
tunneling density of states (DOS) measurements we show that, although the
magnitude of the AHE initially increases with increasing disorder, it eventually saturates and becomes disorder independent once the film resistance reaches values that are of the
order of the quantum resistance
$R_Q=h/e^2$.  Spin-resolved tunneling DOS spectra show
that the conduction electron polarization of the films remains constant throughout the
region of study, and that the saturation behavior is associated with a rapid attenuation of states near the
Fermi energy as $R\rightarrow R_Q$.

It is now well-established that, in the weak-localization limit, $e$-$e$ interactions produce
a perturbative logarithmic depletion of states near the Fermi energy.  As the sheet
resistance of the film is increased the magnitude of the depletion
region grows.  Finally, as one approaches $R_Q$ the effects of $e$-$e$ correlations on the DOS
are no longer perturbative and, in fact, a correlation gap
begins to open in the DOS spectrum \cite{Butko}. In this regime the transport
usually exhibits a modified variable-range hopping form, and macroscopic phases that are nominally robust to disorder begin to become affected.  For instance, if the system is a superconductor, then the superconducting phase is usually lost as $R\rightarrow h/4e^2$ \cite{SIrev}.

Altshuler and coworkers \cite{Altshuler}
first showed that the ordinary Hall conductivity exhibits coherent
scattering corrections but no $e$-$e$ interaction corrections.
Extensions of this work to the AHE also predicted that there would be no
interaction corrections to the AHE conductivity in the weak disorder
limit; which was, indeed, verified in measurements of quenched
condensed Fe films in the early 1990's \cite{Bergmann,Langenfeld}.
However, more recent
measurements on Fe/Si multilayers and high-resistance Fe films have
suggested that disorder-enhanced $e$-$e$ correlations are reflected
in the scaling behavior of the AHE conductivity \cite{Lin,Mitra}.
Though it is now
evident that quantum correlations emerge as the disorder is
increased beyond the weak limit, the current state of understanding
does not shed much light on the influences dimensionality, microscopic
morphology, and scattering mechanisms have on the manifestations of a
correlation-modified AHE.   Furthermore, previous studies have
primarily reported scaling behavior in which the film resistance was
varied by changing the temperature. Here, in contrast, we keep
temperature fixed and vary the thickness/disorder of the films.  AHE measurements in ultra-thin CNi$_3$
films are presented as a function of sheet resistance in the range of $R_{Q}/100$ to $R_{Q}$
at a constant temperature of 2 K.  In addition, tunneling DOS
spectroscopy is used to measure both the strength of the
$e$-$e$ correlation effects and the electron polarization in the films.

Thin films of the metastable intermetallic CNi$_{3}$ were deposited
onto liquid nitrogen cooled, fire-polished glass substrates
by electron-beam evaporation of CNi$_{3}$ targets.  Stencil masks
were used to form either a Hall geometry ($1.5$ mm x $4.5$ mm) or a film pattern more
appropriate for planar tunneling measurements. The films formed a
dense, homogenous C-Ni base which showed no sign of granularity down
to the 1~nm scale using atomic force microscopy.  Some carbon, in the form of multiwall carbon nanotubes, was precipitated out during
the quench condensation of the vapor, but their density was too low to affect the transport \cite{Young}.  Films in the
thickness range of 2 to 10~nm were metallic in appearance, though
partially transparent, and adhered extremely well to the glass
substrates. Resistance and Hall measurements were made in Quantum
Design PPMS using a standard four-wire dc I-V method. Probe currents
of 1~mA and 1~$\mu$A were used in Hall effect and transport
measurements, respectively.  The samples were vapor cooled down to 2~K
in magnetic fields $\pm$9~T via the PPMS. The details of
preparation of superconducting Al tunneling counter-electrodes on the
films have been described elsewhere \cite{Xiong}.  The tunneling
data was taken in a dilution refrigerator equipped with an {\it in
situ} mechanical rotator and a 9~T superconducting solenoid.

\begin{figure} \centering
\includegraphics[width=.5\textwidth]{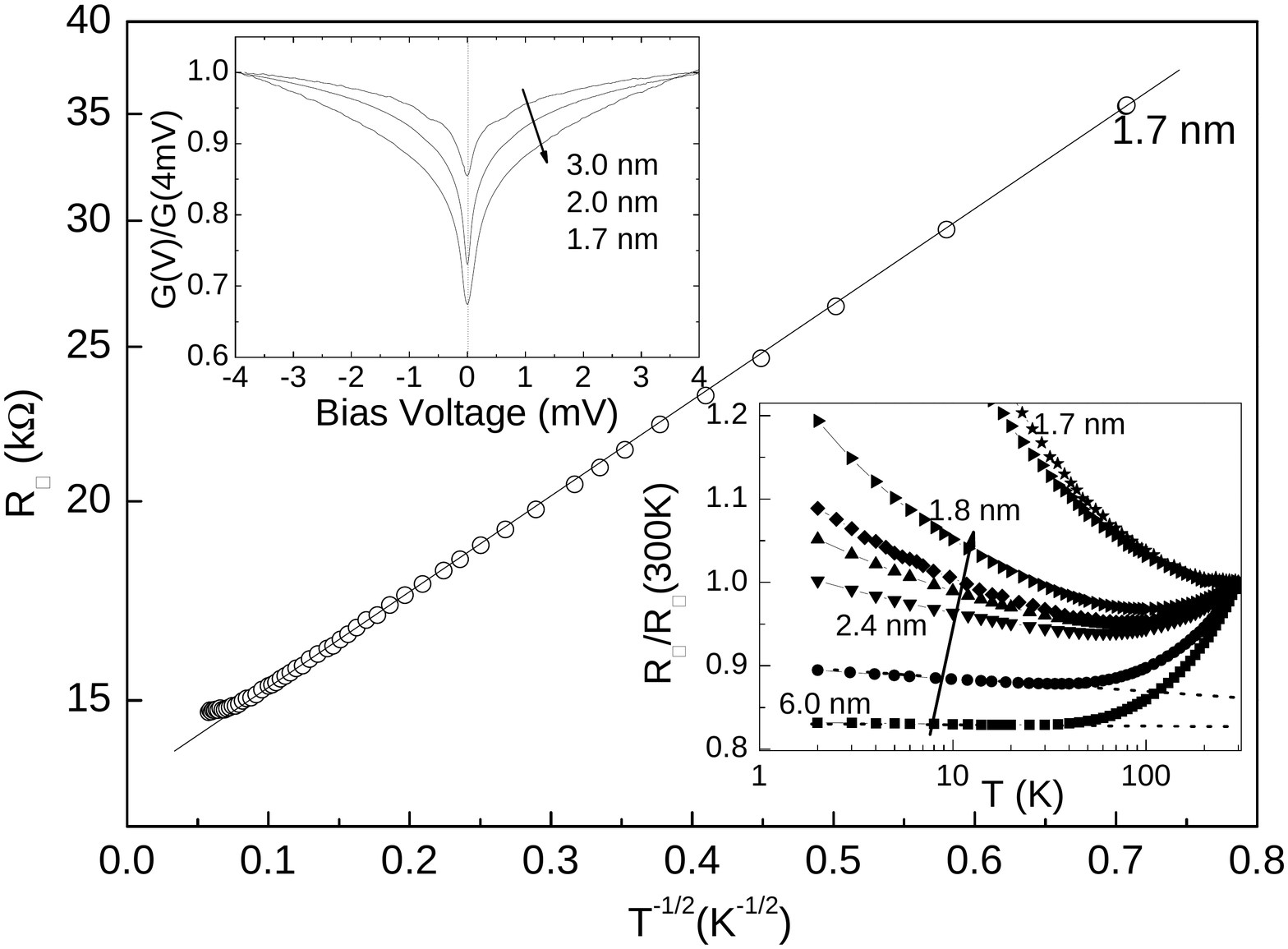}
\caption{The main panel shows a semi-log plot of the sheet resistance of
1.7 nm-thick CNi$_3$ film as a function of $T^{-1/2}$.  The solid
line is a fit to Eq.\ \ref{VRH} .  Upper inset: Electron tunneling
conductance at 100 mK as a function of bias voltage for three CNi$_3$
films of varying thickness.  Lower inset: The normalized sheet resistance as a
function of $\ln T$ for CNi$_3$ films of thicknesses 6.0, 4.0, 2.4,
2.1, 1.9, 1.8, 1.75, and 1.7 nm.}
\label{R-T}
\end{figure}

\begin{figure}
\includegraphics[width=0.5\textwidth]{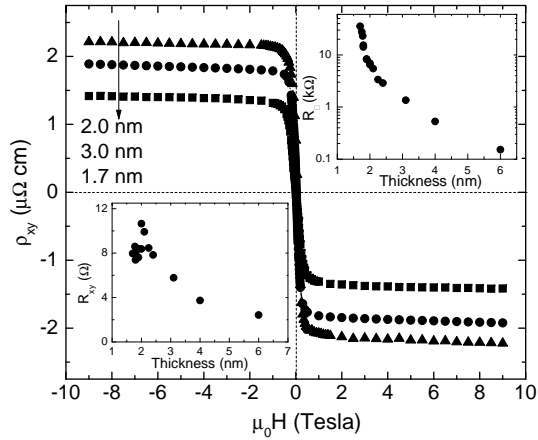}
\caption{The 2~K Hall resistivity of three films with the same
thicknesses as those in the upper inset of Fig.~\ref{R-T}.  Upper inset: CNi$_3$ sheet resistance as a function of film thickness at 2 K. Lower inset: Anomalous Hall resistance as a function of film thickness at 2 K.}
\label{AHE}
\end{figure}

The lower inset of Fig.\ \ref{R-T} shows the sheet resistance $R$ of
several CNi$_{3}$ films of varying thickness as a function of $\ln
T$. Though the 6 and 4~nm films exhibit metallic behavior at high
temperatures, a logarithmic temperature dependence emerges at low
$T$, as highlighted by the dashed lines in the inset.  This behavior
is a hallmark of two-dimensional weak-localization and is clearly
evident below 20~K \cite{Lee}.  However, the film disorder increases rapidly
with decreasing thickness (see Fig.~\ref{AHE} inset), and the resistance of the
thinner films increases faster than $\ln T$.   The upper inset in
Fig.~\ref{R-T} is the tunnel junction conductance of three CNi$_3$
films of varying thickness.  These data were taken at 100~mK in a 7~T
magnetic field oriented parallel to the film surface.   The field
was chosen to be above the parallel critical field of the Al
counter-electrode, and, at this low temperature, the tunneling
conductance was simply proportional to the DOS
of the film \cite{Tinkham}.  The dip in the DOS spectrum at $V = 0$
reflects the
suppression of states near the Fermi energy as a result of $e$-$e$
interactions and is commonly referred to as the zero bias anomaly
\cite{Altshuler,Butko}.
In 2D the anomaly varies logarithmically with energy, $\Delta
G/G\sim \ln(V)$, in the weak disorder limit. The 3~nm spectrum in
the inset figure does, indeed, vary as $\ln V$ and agrees well with that measured in thin Be films \cite{Butko}. However, as the
disorder is increased, the strength of the anomaly grows and
eventually becomes algebraic in energy, $\Delta G/G\sim V^\alpha$.
Deep in the strong localization limit, $R\gg R_Q$, the previously
perturbative anomaly becomes a full-blown Coulomb gap with
$\alpha=1$ \cite{Butko}.  In this regime the transport is expected to be of a
modified variable range hopping form \cite{Shklovskii},
\begin{equation}
\label{VRH}
R(T)=R_0\exp(T_0/T)^{1/2},
\end{equation}
where $R_0$ is a constant of the order of $h/2e^2$ and $T_0$ is the
correlation energy.   The main panel of Fig.~\ref{R-T} shows a
semi-log plot of sheet resistance as a function of $T^{-1/2}$
for a 1.7~nm thick film, which is the thinnest used in this study.
Fitting the data to Eq.~\ref{VRH} gives $T_0=1.9$ K.  Interestingly, the variable range hopping behavior extends to temperatures well above $T_0$ \cite{VRH-T} and is observed in a region where the Coulomb gap, as measured by tunneling, is not fully developed. Nevertheless, the tunneling data show a non-perturbative depletion of states in the thinnest films that is consistent with what has been reported in critically disordered Be films \cite{Butko}.  Given this clear signature of the onset of the correlated insulator regime, we now turn to the evolution of the AHE.

In Fig.~\ref{AHE} we plot the 2~K Hall resistivity,
$\rho_{xy}=R_e\mu_oM$, as a function of applied perpendicular field for three
CNi$_3$ films of the same thicknesses as used in the upper inset of
Fig.~\ref{R-T}.  The step-like structure at low fields
($\mu_oH<1$ T) represents the AHE, and the shallow-sloped higher
field data represent the ordinary Hall effect.  We note that, in the
weak- to moderately-disordered regime, the magnitude of the anomalous
Hall resistivity is expected to increase as a super-linear power of
the sheet resistance $\rho_{xy}\sim R^\beta$ \cite{Nagaosa,Bergmann}.
However, the data in Fig.~\ref{AHE} show that, in the thinnest films, $\rho_{xy}$ is neither
monotonic in film thickness nor sheet resistance.

\begin{figure}
\includegraphics[width=0.5\textwidth]{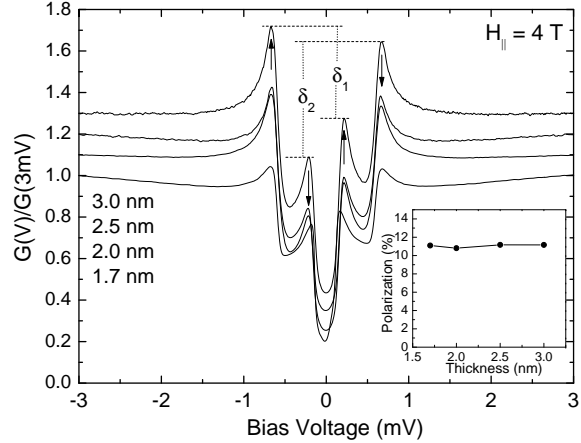}
\caption{The normalized tunneling conductance as a function of bias voltage for
Al-AlO$_x$-CNi$_3$ tunnel junctions on different thickness CNi$_3$
films in a 4 T parallel magnetic field at 100~mK. The curves have been shifted vertically
for clarity.  Inset: The corresponding polarizations of the 1.7, 2.0, 2.5 and 3.0 nm CNi$_3$ films.}
\label{Zeeman}
\end{figure}

In principle, the local maximum in $\rho_{xy}$ could be a consequence
of a thickness-dependent magnetization.  A particular concern is
that the magnitude of the magnetization is suppressed in films with
$t<2$ nm, which, of course, would be reflected in a smaller AHE
\cite{Bergmann2,Kurzweil}. To
address this issue we have employed Tedrow and Meservey's spin
polarized tunneling technique \cite{Tedrow} to directly measure the
electron polarization in the CNi$_3$ films.  The technique exploits
the fact that, when a magnetic field is applied in the plane of a
thin superconductor, in our case the Al counter-electrode, the BCS
DOS spectrum is Zeeman-split into well resolved spin
sub-bands.   Figure~\ref{Zeeman} shows the normalized tunneling
conductance vs. bias voltage of Al-AlO$_x$-CNi$_3$ tunnel junctions
on different thickness CNi$_3$ films in a 4~T parallel magnetic
field.   The arrows denote the respective spin assignments of the
occupied and unoccupied sub-band peaks.  When tunneling into a
paramagnetic metal the peaks are symmetrically positioned about $V=0$.
However, as is clearly evident in the figure, the peak heights are
asymmetric, reflecting the unequal spin populations in the
ferromagnetic CNi$_3$.  The electron polarization $P$ is obtained by
measuring the relative heights of the peaks \cite{Tedrow,Xiong},
\begin{equation}\label{Polarization}
P=\left|\frac{\delta_{1}-\delta_{2}}{\delta_{1}+\delta_{2}}\right|,
\end{equation}
where the peak height differences $\delta_{1}$ and
$\delta_{2}$ are defined in Fig.~\ref{Zeeman}.  The electron
polarization of the CNi$_3$ films with different thickness were are all
about 11$\%$. In the inset of Fig.~\ref{Zeeman} we plot $P$ as a
function of film thickness. Although polarization is not equivalent
to magnetization, direct measurements of $P$ in a variety of Ni alloys \cite{Meservey} has shown that the $P$ accurately tracks the saturation magnetization across a wide range in $M$.  Thus the data in Fig.~\ref{Zeeman} strongly suggest
that the magnetization of CNi$_3$ films remains unchanged in the
thickness range of 1.7 to 3~nm.  Thus the non-monotonic behavior in Fig.~\ref{AHE} is due to quantum
corrections to the anomalous Hall coefficient $R_e$.

\begin{figure}
\centering
\includegraphics[width=0.5\textwidth]{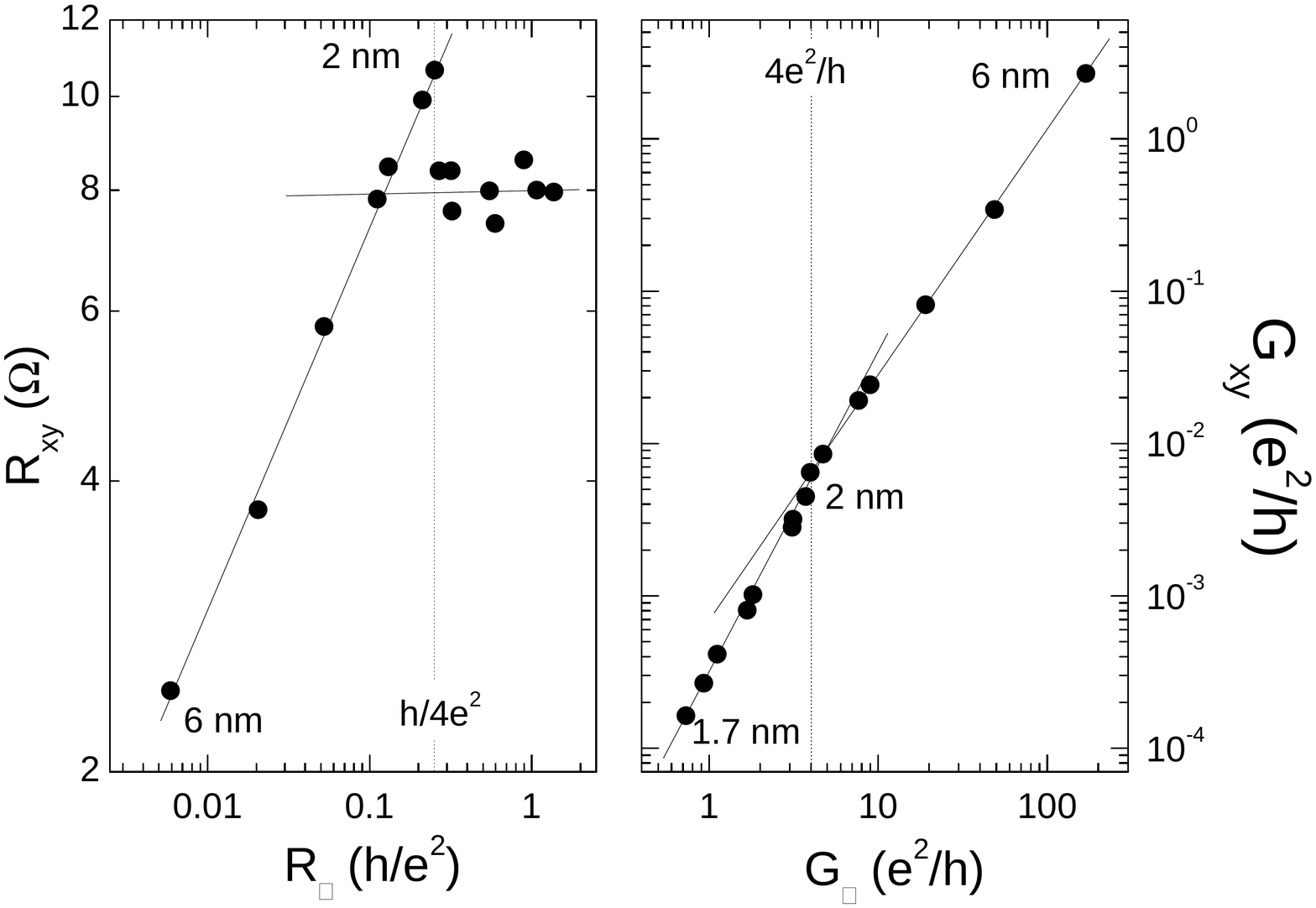}
\caption{Left panel: Log-log plot of the anomalous Hall resistance of
CNi$_3$ films as a function of the sheet resistance at 2~K.  The
solid lines provide a guide to the eye.  Right panel:  Log-log
plot of the anomalous Hall conductance as a function of the sheet
conductance at 2~K.  The solid lines represent separate power law
fits to the low and high conductance data,
giving exponents of $\varphi=2.0$ and $1.6$, respectively.}
\label{Scaling}
\end{figure}

The evolution of the AHE with increasing disorder is
shown in the log-log plots of Fig.~\ref{Scaling}.  Because of the 2D nature of the transport, we present the data in terms of longitudinal sheet resistances and conductances, along with the corresponding Hall resistances and conductances.   In the left
panel we show the anomalous Hall resistance as a function of
sheet resistance at 2~K.    As expected $R_{xy}$ initially increases
with increasing $R$, but near $R=h/(4e^2)$ the Hall
resistance abruptly saturates.  The saturation threshold occurs at a
film thickness $t\sim2$ nm.  In the right panel of Fig.~\ref{Scaling}
 we plot the anomalous Hall conductance $G_{xy}\approx
R_{xy}/R^2$ as a function of the sheet conductance.  Note that there
is a kink in the curve near $G=4e^2/h$ where the scaling exponent
changes from 1.6 to 2.0 as the film thickness is lowered below the 2
nm threshold.  Recent numerical work on the AHE in ferromagnetic
transition metals suggests that the Hall conductivity in the
moderately disordered regime is dominated by intrinsic mechanisms
\cite {Sangiao,Kontani,Onoda} as opposed to extrinsic scattering
processes that are prominent in very low disorder systems
\cite{Nagaosa}.  In the moderately disordered, weak localization
regime of $10^{4}~$S/cm$\lesssim\sigma_{xx}\lesssim10^{6}~$S/cm the
AHE is dissipationless with $\sigma_{xy}\sim R^0$ .  However in the
``dirty metal'' regime,
$10^{3}~$S/cm$\lesssim\sigma_{xx}\lesssim10^{4}~$S/cm the Hall
conductivity scales as $\sigma_{xy}\sim R^\varphi$, where $\varphi$
ranges between 1.6 and 2.0, depending on the details of the
calculation \cite{Kontani,Onoda}.  We note that the
low-temperature conductivity of our films varies from
$\sim10^{4}~$S/cm to $\sim10^{3}~$S/cm in the thickness range of 6
to 2~nm.  Thus, we believe that the data in the 2 - 6~nm range of
Fig.~\ref{Scaling} is, in fact, in the ``dirty metal'' regime, with
a corresponding exponent $\varphi=1.6$.  This is in good agreement
with what has been reported in a wide variety of itinerant
ferromagnets of similar disorder \cite{Sangiao,Fernandez-Pacheco,Miyasato}. The knee in the scaling plot probably reflects the fact that the
films cross over from logarithmically localized transport to modified
variable-range hopping transport in the vicinity of the quantum
resistance.  This crossover is, of course, commensurate with the
rapid depletion of states near Fermi energy, see upper inset of Fig.\ \ref{R-T}, which is a precursor to the emergence of the Coulomb gap \cite{Butko}. Recently Nagaosa {\it et al}.\ \cite{Nagaosa} have
suggested that the low-temperature phase diagram of the AHE should include
a phase boundary near the Mott-Anderson critical region. They estimate the zero-temperature
boundary should be in the vicinity of $\sigma=10^3~\Omega^{-1}$cm$^{-1}$, which roughly
corresponds to the separatrix in the right panel of Fig.\ \ref{Scaling}.

In summary, we show that the anomalous Hall resistance saturates in
homogeneous CNi$_3$ films with sheet resistance near the quantum resistance, although the electron polarization remains unchanged.   We believe that the saturation is associated with the crossover from the
weak localization regime to that of a 2D correlated insulator.  The
crossover is also clearly evident in the scaling behavior of the Hall
conductivity as well as in the tunneling density of states.

\begin{acknowledgments}
We thank John DiTusa and Ilya Vekhter for enlightening discussions. This work was supported by the DOE under Grant No.\ DE-FG02-07ER46420.
\end{acknowledgments}

\end{document}